# Creation of flexible spin-caloritronic material with giant transverse thermoelectric conversion by nanostructure engineering

Ravi Gautam[1], Takamasa Hirai[1], Abdulkareem Alasli[2], Hosei Nagano[2], Tadakatsu Ohkubo[1], Ken-ichi Uchida[1*], Hossein Sepehri-Amin[1*]

[1]National Institute for Materials Science, Tsukuba, 305-0047, Japan

[2]Department of Mechanical Systems Engineering, Nagoya University, Nagoya 464-8601, Japan

**Functional materials such as magnetic, thermoelectric, and battery materials have been revolutionized through nanostructure engineering. However, spin caloritronics, an advancing field based on spintronics and thermoelectrics with fundamental physics studies, has focused only on uniform materials without complex microstructures. Here, we show how nanostructure engineering enables transforming simple magnetic alloys into spin-caloritronic materials displaying significantly large transverse thermoelectric conversion properties. The anomalous Nernst effect (ANE), a promising transverse thermoelectric phenomenon for energy harvesting and heat sensing, has been challenging to utilize due to the scarcity of materials with large anomalous Nernst coefficients. We demonstrate a remarkable improvement in the anomalous Nernst coefficients in flexible Fe-based amorphous materials through nanostructural engineering, without altering their composition. This surpasses all reported amorphous alloys and is comparable to single crystals showing large ANE. The enhancement is attributed to Cu nano-clustering, facilitating efficient transverse thermoelectric conversion. This discovery advances the materials science of spin caloritronics, opening new avenues for designing high-performance transverse thermoelectric devices for practical applications.**

*Corresponding authors. *Email address:* h.sepehriamin@nims.go.jp (H. Sepehri-Amin) and uchida.kenichi@nims.go.jp (K. Uchida).

Thermoelectric devices have emerged as a promising technology for generating electricity by scavenging waste heat, representing a vital step towards achieving a more sustainable future[1–3]. Current technology, which relies on the Seebeck effect, employs a complex module structure with inadequate thermal insulation[4,5]. To overcome this problem, research interest in transverse thermoelectric conversion such as the anomalous Nernst effect (ANE) has been raised[3,6]. ANE can convert a temperature gradient into an orthogonal voltage in magnetic materials (Fig. 1) and is the Onsager reciprocal of the anomalous Ettingshausen effect (AEE)[3,5,7–12]. ANE is one of the hottest topics in spin caloritronics[5,6,13] because it enables the design of thermoelectric devices with a simple lateral structure, convenient scalability, and easy fabrication. Additionally, its transverse geometry facilitates efficient and flexible coverage of curved heat sources, making it ideal for harvesting thermal energy from large-area heat sources. Compared to conventional Seebeck-effect-based devices, ANE-based devices require fewer manufacturing processes, exhibit lower contact resistance, and may offer superior thermoelectric conversion performance[2,14]. This makes ANE a promising path for the development of next-generation thermoelectric devices with potential applications ranging from thermal management technologies to heat flux sensors[2]. However, further improvement of the thermopower of ANE is needed, which is the current bottleneck for its practical applications. As the demand for modern devices continues to rise, achieving a significant transverse thermoelectric response and understanding the underlying mechanisms become crucial for optimizing efficiency and power generation capabilities. To accomplish this, a novel approach is required to design low-cost and flexible materials that exhibit large ANE, thereby advancing its use in cutting-edge thermoelectric devices.

The discovery of remarkable ANE in magnetic topological materials has triggered a surge of interest, and significant efforts have been directed toward enhancing ANE through materials research. For example, several magnetic materials such as Heusler compounds[7–9,15–19], ferromagnetic binary alloys[10,12,27,14,20–26], Weyl antiferromagnet[28–30], and permanent magnets[11,31–33] have been extensively investigated. However, only a few of these materials have displayed the anomalous Nernst coefficient $S_{ANE}$, transverse thermopower due to ANE, surpassing 1 µVK$^{-1}$, with Co-based Heusler alloys demonstrating particularly remarkable values of 6-8 µVK$^{-1}$ at room temperature[8]. Additionally, while most research has predominantly focused on crystalline materials, there has been a recent emergence of studies exploring amorphous materials[33–39]. Iron (α-Fe) is known as the most abundant ferromagnetic element, but it exhibits tiny ANE[10,12,20]. Thus, it would be favorable to develop a versatile Fe-based bulk material with large ANE to obtain highly efficient thermoelectric devices. So far, particular emphasis has been directed on material design and development to increase the ANE via manipulating the Berry curvature near the Fermi level[11,17,19,22,28]. However, little attention has been paid to how microstructure engineering at different length scales can influence the transport properties and ANE. This can potentially unveil new strategies to enhance ANE, thus advancing our understanding and application of this effect in thermoelectric and energy conversion devices.

In this study, we demonstrate how nanostructural engineering of magnetic materials can significantly improve the transverse thermoelectric conversion properties. We present a schematic representation of our strategies in Fig. 1. To confirm the validity of our approach, we demonstrated this concept by investigating ANE in Fe-based amorphous materials, commercially known as "Nanomet"[40], by tailoring their nanostructure. Nanomet is a renowned soft magnetic material that contains a high amount of Fe (~ 83- 85 at%) and cost-effective alloying elements (i.e., Cu, B, P, and Si). This material is widely employed in high-frequency power electronic devices, such as inductors and transformers, owing to its excellent soft magnetic properties. Surprisingly, we found that, although the as-prepared Fe-based amorphous alloy initially exhibited low ANE, optimal annealing dramatically increased $S_{ANE}$ despite the same average composition, owning to the formation of Cu nanoclusters with a large volume fraction. Notably, this material exhibits largest $S_{ANE}$ among amorphous alloys[33–39]. The study involved four steps: (a) starting with the as-quenched Nanomet samples with small ANE, (b) forming Cu clusters in the amorphous samples, (c) optimizing the density of Cu-rich clusters embedded in an amorphous matrix with a significant enhancement of ANE, and (d) transforming the samples into crystallized Nanomet ribbons with small ANE with further increasing annealing temperature. These strategies were devised to understand the influence of nanostructure on ANE/AEE and provide insights into the thermoelectric properties of the studied materials. Our work demonstrates that nano-sized Cu-rich clusters in an amorphous matrix of Fe-based soft magnetic materials (Nanomet) lead to a significant 70% enhancement of $S_{ANE}$ without altering the material chemistry. This strategy resulted in the figure of merit for ANE comparable to the record-high value in a Heusler ferromagnet at room temperature[7]. The developed materials are easily mass-produced at a low cost, and their mechanical flexibility enables the fabrication of large-area ANE and AEE devices with various shapes. This study and proof of concept thus open new avenues for enhancing the ANE/AEE properties of materials in general through a nanostructure engineering approach and discusses the correlation between the nanostructure and its transport properties.

**Nanostructure engineering of amorphous Fe-based ribbons**

We have fabricated an 18 ± 2 μm thin and several meters long Nanomet ribbons using the melt-spinning technique which brings great benefits for low manufacturing and raw material costs. Differential scanning calorimetry analysis of the Nanomet ribbons showed the crystallization of α-Fe(Si) and Fe-compound phases at respective onset temperatures of $T_{x1}$ = 661 K and $T_{x2}$ = 806 K (Supplementary Fig. 1). Phase analysis of the as-quenched (as-spun) sample by X-ray diffraction (XRD) exhibits a broad halo peak indicating the presence of a fully amorphous phase (Supplementary Fig. 2) which was further confirmed by the electron diffraction pattern obtained using transmission electron microscopy (TEM). Figure 2a displays a high-resolution bright field (BF) scanning (S)TEM image obtained from the as-quenched sample. The uniform supersaturated solid solution with an amorphous structure was observed in this sample. The amorphous ribbons were annealed at various isothermal temperatures (573, 623, 643, 653, 673, and 723 K) for only 3 min to achieve different microstructural features and investigate their correlations with ANE. The microstructure of the annealed samples was studied by high resolution BF-STEM and electron beam diffraction analysis as shown in Fig. 2b-e. Annealing at the temperatures below 653 K does not change the amorphous microstructure, while annealing at the temperatures above 673 K leads to the crystallization of the α-Fe(Si) phase (Supplementary Fig. 2 and Fig. 3). The elements P, B and C present in the Nanomet alloy composition, are known to be amorphous forming elements[41] and this is the main reason for the formation of an amorphous phase in the as-quenched ribbons, which inhibits the nucleation of α-Fe(Si). Note that the obtained amorphous structure provides excellent flexibility to the ribbons developed. As the temperature was increased beyond the crystallization points, there was a notable reduction in the presence of amorphous phases, accompanied by an increase in the nucleation of the α-Fe(Si) phase. Subsequent annealing at 723 K resulted in a significant volume fraction of the α-Fe(Si) phase with a larger grain size. Detailed examination through high-resolution TEM images revealed the distinct separation of the α-Fe(Si) crystals from each other by the residual amorphous matrix. Moreover, the XRD analysis provided evidence for the appearance of additional diffraction peaks corresponding to the crystallization of the Fe-P and Fe-B compounds.

**Nano-scale elemental analysis of Fe-based ribbons**

A question is if there is any nano-scale elemental fluctuation in an amorphous matrix before and/or after crystallization of α-Fe(Si) crystals. To answer this question, we conducted atom probe tomography (APT). 3D elemental maps of Fe, P, B, and Cu are shown in Fig. 3a-d. The illustrated maps are ~10 nm thin sliced from whole data for a better visualization of the elemental distributions. A uniform distribution of all elements was observed in the as-quenched sample as shown in Fig. 3a, indicating its chemically homogeneous solid solution. As shown in Fig. 3b, a heterogeneity in the distribution of Cu atoms in the microstructure of the annealed sample at 623 K was detected. This observation strongly implies the occurrence of Cu clustering, indicated by the segregation of Cu atoms. As the annealing temperature was raised to 653 K, Cu clustering became more visible as demonstrated in Fig. 3c, with an increase in both the size and number density of nano-sized Cu-rich clusters. Notably, no crystallization was observed up to 653 K (Fig. 2d), indicating the presence of Cu clusters dispersed within the amorphous matrix. However, due to the absence of discernible fringe contrast in the nanobeam TEM analysis, the specific structure of these clusters could not be identified. The clustering phenomenon of Cu in Fe-based material can be elucidated by considering the markedly positive enthalpy of mixing exhibited by this system. The low solubility of Cu in Fe leads to the formation of Cu-enriched clusters within the amorphous phase, preceding the nucleation of primary α-Fe crystals. These clusters then act as heterogeneous nucleation sites for the primary α-Fe(Si) crystals[42]. This requires phase separation within the amorphous phase, which can be seen in Fig. 3d where Cu precipitates were observed in direct contact with α-Fe(Si) grains in a sample annealed at 723 K. The observed increase in cluster size in conjunction with a decrease in number density suggests the occurrence of the classic Ostwald ripening phenomenon. Thus, the microstructure transformation in the Nanomet melt-spun ribbon occurs through a series of three distinct stages: (1) an initial amorphous phase, (2) an amorphous phase with the formation of clusters, and (3) nucleation of an α-Fe(Si) phase at the heterogeneous sites of clusters trailed by subsequent grain growth, reduction in the amorphous phase and the appearance of additional Fe-B compound phases.

**Thermal transport properties of nanostructure engineered Fe-based ribbons**

The thermal transport properties of the nanostructure engineered Nanomet ribbons were investigated. Figure 4a displays the thermal conductivity $\kappa$ of the annealed samples, estimated via $\kappa = \rho C_p D$ using the thermal diffusivity $D$, specific heat $C_p$, and material density $\rho$ (Supplementary Figs. 4-7). $D$ was measured using a laser-spot-periodic-heating method based on lock-in thermography (LIT) (see Methods and Supplementary Figs. 4-6)[43]. No significant change in the $D$ values was observed for the samples annealed below the onset of crystallization temperature ($T_{x1}$ = 661 K) and the values were found to be in a range of 2.2 to 2.6 × $10^{-6}$ $m^2s^{-1}$. A steep rise in $D$ was observed as the crystallization of α-Fe(Si) started in the amorphous matrix for samples annealed at 673 K and 723 K, with an average value of 3.5 × $10^{-6}$ $m^2s^{-1}$ and 4.1 × $10^{-6}$ $m^2s^{-1}$, respectively (Supplementary Fig. 5). $C_p$ did not reveal any discernible patterns, with values ranging between 0.47 and 0.56 $Jg^{-1}K^{-1}$ (Supplementary Fig. 7), closely resembling the specific heat of pure Fe. Figure 4a shows that $\kappa$ increases with the annealing temperature, with a sudden rise in its value after 673 K. Thus, the observed variations in $\kappa$ are primarily attributed to the changes in $D$, while $C_p$ remains nearly constant throughout the annealing process. The electrical conductivity $\sigma$ of the annealed samples was also studied to estimate the contribution of phonons and electrons in $\kappa$ through the Wiedemann-Franz law[44]. Figure 4b indicates that $\sigma$ exhibits a similar trend to $\kappa$ with a substantial increase beyond the crystallization temperature. The electron and phonon thermal conductivities were respectively estimated as $\kappa_e = \sigma LT$ and $\kappa_p = \kappa - \kappa_e$, with $L$ being the Lorenz number (2.44 × $10^{-8}$ $W\Omega K^{-2}$) and $T$ the absolute temperature (300 K) (Supplementary Fig. 8). Interestingly, phonons and electron have comparable contribution to the thermal conductivity for the samples annealed below 643 K. However, a slight dominance of the phonon contribution was observed for the sample containing a high-density of Cu-clusters (annealed at 653 K) and for the samples annealed beyond the crystallization temperature. Hence, $\kappa$ and $\sigma$ behaviours can be attributed to the well-ordered atomic structure in the crystalline materials, which facilitates efficient phonon/electron transport and thus leads to higher conductivities. In contrast, the amorphous phase, characterized by structural disorder and more scattering sites, hinders phonon/electron transport, and resulting in the smaller $\kappa_e$ and $\kappa_p$ values. It is worth noting that in our samples, the formation of clusters increases $\kappa$ due to the presence of Cu-rich clusters in the amorphous matrix, while the presence of clusters can also affect these properties by introducing additional scattering sites for both phonons and electrons. Therefore, the formation of a high-density Cu-clusters in the amorphous matrix facilitates the electron and phonon transport by reducing the volume fraction of the disordered structure. In the future, the phonon engineering approach can be implemented to fine-tune the transport properties[45].

**Large anomalous Nernst effects in nanostructure engineered Fe-based amorphous ribbons**

To investigate the impact of microstructure tailoring on the transverse thermoelectric properties, we studied AEE using the thermoelectric imaging technique based on LIT method. This technique allowed us to examine the spatial distribution of temperature modulation and symmetry of AEE. Steady-state AEE signals were analyzed to quantitatively estimate the anomalous Ettingshausen coefficient ($\Pi_{AEE}$) at room temperature (300 K) (see Fig. 4c, Methods, and Supplementary Fig. 9). The corresponding $S_{ANE}$ values of the samples were determined through the Onsager reciprocal relation: $S_{ANE} = \Pi_{AEE}/T$, also presented in Fig. 4c. The magnitude of $S_{ANE}$ for the as-quenched sample was estimated to be 2.2 μVK$^{-1}$. This value is already higher than that for the other polycrystalline Fe-based alloys and an order of magnitude larger than that for pure Fe[2,12]. During annealing, the $S_{ANE}$ value remained relatively stable up to 623 K. It increased to 2.7 μVK$^{-1}$ for the sample annealed at 643 K and surprisingly shoots up to a remarkably higher value of 3.7 μVK$^{-1}$ for the sample annealed at 653 K, which is 70% higher than that for the as-quenched sample. However, when the annealing temperature exceeds 673 K and the crystallization of α-Fe(Si) proceeds, the $S_{AEE}$ values decrease. The substantial increase in $S_{ANE}$ was only observed for the sample with a high-density of Cu-enriched clusters within the amorphous matrix. This could be attributed to the interfacial spin-orbit interaction at the boundaries between the Fe-based amorphous matrix and the Cu-clusters[46]. In addition, the crystallization of α-Fe(Si) leads to a compositional change in the amorphous phase, which could be optimized in the fully amorphous samples with large ANE. Although the formation of Cu-clusters in the amorphous matrix resulted in the high $S_{ANE}$ value, theoretical studies are desirable to elucidate the microscopic origin[22].

In Fig. 5a and 5b, we respectively compared the |$S_{ANE}$| and power factor (PF) $\sigma S_{ANE}^2$ values for various spin-caloritronic amorphous materials reported in the literature with the Fe-based amorphous materials developed in this study. Notably, the Nanomet sample annealed at 653 K exhibits a significantly large $S_{ANE}$ value (~3.7 μVK$^{-1}$) and a high PF (~7.7 μWm$^{-1}$K$^{-2}$) at room temperature, which are larger than the values observed in all of the conventional spin-caloritronic amorphous materials and comparable to the single-crystalline materials reported so far[2,8,10,29]. Furthermore, it is important to highlight that most of the bulk polycrystalline materials exhibit |$S_{ANE}$| values of <1 μVK$^{-1}$, while only a few materials, such as the SmCo$_5$ and Co$_2$MnGa systems, possess $S_{ANE}$ of >3 μVK$^{-1}$ at room temperature[8,12,29]. However, both are expensive due to the existence of Co or Sm and are not flexible. Therefore, it is worth noting that our low-cost and flexible material demonstrates the high $S_{ANE}$ value at room temperature, achieved through nanostructure engineering. Figure 5c highlights the mechanical flexibility of the Nanomet materials below the onset crystallization temperature. This figure vividly demonstrates their remarkable ability to be easily curved into various shapes, confirming their suitability for practical applications that demand flexibility. This nanostructure engineering approach enables the enhancement of $S_{ANE}$ within the same materials, leading to elevated levels of performance.

To assess the effectiveness of the Nanomet ribbons as spin-caloritronic materials for ANE/AEE applications, we estimated the dimensionless figure of merit for ANE through the equation[32]: $Z_{ANE}T = \frac{S_{ANE}^2 \sigma}{\kappa} T$. As shown in Fig. 4d, the maximum $Z_{ANE}T$ value was found to be 2.2 × 10$^{-4}$ at $T$ = 300 K for the sample annealed at 653 K, which is three orders of magnitude larger than that for pure Fe and comparable to the record-high $S_{ANE}$ values at room temperature for the SmCo$_5$-type permanent magnets[31,32] and Co$_2$MnGa Heusler ferromagnet[7,8]. This indicates that the Nanomet ribbons with embedded nano-sized Cu-clusters in an amorphous matrix can offer an excellent combination of microstructure features for enhancing the transverse thermoelectric conversion properties. From Fig. 4, it can be inferred that the large value of $S_{ANE}$ is the dominant factor contributing to the improvement of $Z_{ANE}T$. However, it is still insufficient for practical applications. Therefore, $S_{ANE}$ and $Z_{ANE}T$ are necessary to be further improved by tailoring the size, distribution, and composition of the clusters and composition of the matrix to achieve the desired functionalities. Larger $Z_{ANE}T$ can also be obtained by reducing $\kappa_p$ through phonon engineering.

In conclusion, we propose and demonstrate a novel approach to design low-cost and flexible spin-caloritronic materials that exhibit large ANE through nanostructure engineering. This approach is of great importance for advancing the use of ANE in thermoelectric devices. These materials have immense potential in the fabrication of highly efficient and flexible energy harvesting and thermal management devices with curvilinear design. This study has established a direct correlation between ANE and engineered nanostructures in a spin-caloritronic material. As a result, we have discovered a new methodology for tailoring ANE using nanoscale clusters embedded in an amorphous matrix microstructure, which is a proof-of-concept for future advancements. The proposed method is applicable to various systems via the development of nanocomposite materials, which can open up a new avenue towards the development of spin-caloritronic materials with giant ANE suitable for practical applications.

**Methods**

**Preparation of Fe-based amorphous alloy.** The master alloy ingot was prepared by melting high-purity elements of Fe (99.99 %), Si (99.99%), Cu (99.9%), B (99.5%), and $Fe_3P$ (99%) using the vacuum induction melting technique under an Ar atmosphere. Subsequently, the amorphous ribbon was produced using the single-roll melt-spinner. The ingot was melted in a quartz tube and then ejected under an Ar atmosphere at a pressure of 0.02 MPa onto a rotating copper (Cu) wheel. The tangential speed of the wheel was set at 35 m/s, while the gap between the quartz tube and the rotating Cu wheel was maintained at 0.2 mm. The process flow chart was optimized to obtain high-quality amorphous ribbons with lengths of several meters, a width of 5 mm, and a thickness ranging from 16 to 20 μm, as shown in Fig. 5c.

For annealing, the samples were placed in quartz tubes and connected to a high vacuum turbo pump to avoid oxidation. To achieve rapid annealing and to prevent undesired microstructural changes during heating, the samples were directly inserted into a preheated tubular furnace. Annealing was performed at different temperatures of 573, 623, 643, 653, 673, and 723 K for a soaking time of 3 min. The temperature of the ribbons was continuously monitored during annealing using a K-type thermometer positioned in close proximity to the ribbons. Notably, a heating rate of nearly 120 K/s was attained, enabling precise temperature control. Further, the ribbons were allowed to cool naturally to room temperature by removing the quartz tube from the furnace.

**Characterization of samples**. The chemical composition was estimated using inductively coupled plasma optical emission spectrometry (ICP-OES) and was found to be $Fe_{84.7}Si_{2.8}B_{3.8}P_{7.8}Cu_{0.7}C_{0.2}$ (at%). This composition is associated with the trademark "Nanomet". The ribbon was cut into rectangular samples with the dimensions of ~ 60 × 5, ~ 10 × 5, and ~ 10 × 1 ± 0.1 mm for $\sigma$ and $D$, and AEE measurements, respectively. The thermal analysis of the as-quenched ribbon was carried out by differential scanning calorimetry (DSC, Rigaku TG8120) at a heating rate of 20 K/min in an Ar atmosphere. The phase analysis of the annealed samples was evaluated by XRD (Rigaku MiniFlex600) with Cr-Kα radiation. Microstructural studies were carried out using a transmission electron microscope (TEM, FEI Titan $G^2$ 80-200 equipped with a probe corrector). The elemental distribution was investigated using atom probe tomography (APT) in the laser mode, utilizing the CAMECA LEAP 5000 XS instrument operating at a repetition rate of 250 kHz. The laser pulse energy was set at 30 pJ, and the experiments were carried out at a base temperature of 30 K, maintaining a constant detection rate of 1.5%. The obtained APT data was subsequently analyzed using CAMECA AP Suit 6.1 software. The TEM specimens and APT tip were prepared by lift-out and annular milling techniques using a dual beam focused ion beam (FEI Helios 5UX).

**Transport properties measurement.** The angular distribution of the in-plane $D$ was measured by means of the LIT method. A schematic of the setup is plotted in Supplementary Fig. 4 comprises an infrared camera, a diode laser, a function generator, and a LIT system. The diode laser emits a modulated laser beam driven by a periodic reference signal at frequency $f$ from the function generator. The modulated beam was focused as a point on the backside of the opaque sample by an optical setup. The laser heat-point in turn induces in-plane heat waves that diffuse radially within the sample. Simultaneously, the camera images the thermal response on the front side of the sample. The LIT system processes the thermal images according to $f$ and outputs the spatial distribution of the lock-in amplitude $A$ and phase $\phi$. $D$ can be estimated from the correlation between $\phi$ and the distance $r$ to the periodic point-heat source within the circular diffusion area. Hence, the angular distribution of $D$ can be obtained by revolving the analysis around the heating point at angles $\theta$ (Supplementary Fig. 4)[47].

$$D = \frac{\pi f}{(d\phi/dr)^2} \tag{1}$$

The measurement was performed with a science-grade LIT system (ELITE, DCG Systems Inc.) and a diode laser of 638 ±1 nm wavelength (LDM637D.300.500, Omicron Inc.). The laser heat spot is focused to a diameter of ~7 μm at a power $P = 10$ mW and modulated at $f = 3$ Hz. $f$ and $P$ are selected by taking into consideration maximizing the signal-to-noise ratio, reducing the heat loss effect[48], and the thermal diffusion length $\Lambda = \sqrt{D/\pi f}$ does not exceed the sample width[47]. The $C_p$ was measured by differential scanning calorimetry (DSC, Rigaku Thermo plus EV02). The $\kappa$ at room temperature was estimated as

$$\kappa = C_p \rho D \tag{2}$$

where $\rho$ is the density of the Nanomet ribbons which is around 7.46 gcm$^{-3}$.

AEE of all the samples was examined also by means of the LIT method[12]. The thermal images of the surface of the samples were obtained by applying a square-wave-modulated periodic charge current with amplitude $J_c$, frequency $f$, and zero offset to the ribbons in the $x$-direction with an applied magnetic field, $\mu_0 H$ along the $y$-direction (Supplementary Fig. 9). The first harmonic response of the detected images was extracted and transformed into $A$ and $\phi$ images through Fourier analysis. Using this methodology, it is possible to isolate and identify the sole effect of thermoelectric effects ($\propto J_c$) without any interference from Joule heating ($\propto J_c^2$)[31,49,50]. In the LIT-based thermoelectric measurements, the $A$ image represents the magnitude of current-induced temperature modulation and the $\phi$ image the sign, that is, $\phi \sim 0°$ (~180°) corresponds to releasing (absorbing) heat, as well as the time delay of the temperature modulation. The LIT measurements were performed by applying $J_c = 1.0$ A at room temperature ($T = 300$ K), $\mu_0 H = \pm 1$ T, and $f = 1.0$-$10.0$ Hz. Since the AEE-induced temperature modulation shows the $H$-odd dependence, we calculated the $H$-odd component of lock-in amplitude $A_{\text{odd}}$ and phase $\phi_{\text{odd}}$ by $A_{\text{odd}} = |A(+H)e^{-i\phi(+H)} - A(-H)e^{-i\phi(-H)}|/2$ and $\phi_{\text{odd}} = -\arg[A(+H)e^{-i\phi(+H)} - A(-H)e^{-i\phi(-H)}]$, where $A(+H)$ [$\phi(+H)$] and $A(-H)$ [$\phi(-H)$] show the $A$ ($\phi$) value measured at $\mu_0 H = +1$ T and $-1$ T, respectively. The AEE-induced temperature modulation at the steady state $A_{\text{odd}}^{\text{steady}}$ was determined by using the magnitude of $A_{\text{odd}}$ signals at $f = 1.0$ Hz because of their $f$-independence in the low $f$ region. The steady-state AEE signals were analyzed to quantitatively estimate the $\Pi_{\text{AEE}}$ values at 300 K using the equation[32]:

$$\Pi_{\text{AEE}} = \pi\kappa|\Delta T_{\text{AEE}}|/4j_c t \tag{3}$$

where $\Delta T_{\text{AEE}}$ represents the temperature change induced by AEE between the top and bottom surfaces of the samples, determined using $|\Delta T_{\text{AEE}}| = 2A_{\text{odd}}^{\text{steady}}$, $j_c$ the charge current density, and $t$ the thickness of the samples. To estimate the figure of merit, $\sigma$ was measured using a four-probe method.

**Acknowledgements**

The authors thank Y. Sakuraba and K. Hono for valuable discussions and M. Isomura, H. Sebata, and N. Kurata for technical supports. This work was supported by JST ERATO "Magnetic Thermal Management Materials" (Grant No. JPMJER2201).

**Author contributions**

R.G and H.S. conceived the idea, planned, and conducted the microstructural characterization. R.G. designed the experiments, prepared the samples, collected, and analysed the data. R.G., H.S., and T.O. conducted atom probe tomography experiments. K.U. conducted AEE measurements and T.H analysed the AEE data. A.A and H.N measured the thermal diffusivity. H.S. and K.U. supervised the study. All authors discussed the results and contributed to preparation and revision of the manuscript.

**Competing interests**

The authors declare no competing financial interests.

**Additional information**

**Correspondence and requests for materials** should be addressed to H.S. or K.U.

**Data availability**

The data that support the findings of this study are available from the corresponding author on reasonable request.

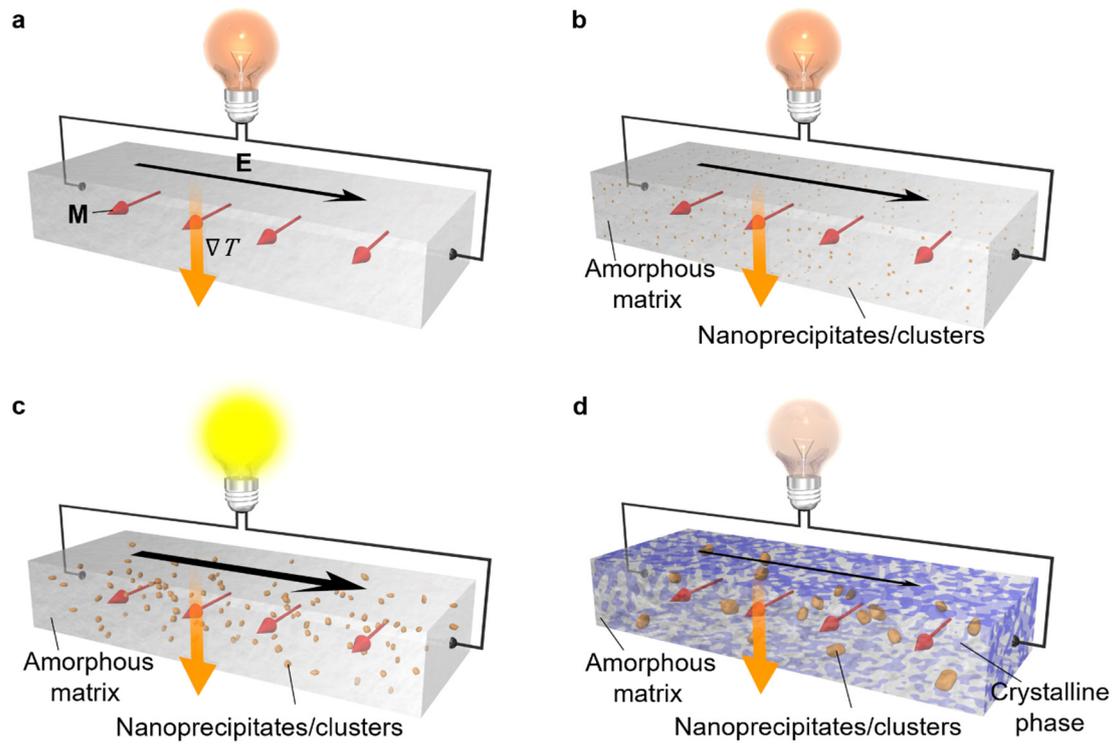

**Fig. 1 | Anomalous Nernst effect in a nanostructure-engineered magnetic material. a-d**, Schematic illustrations of the anomalous Nernst effect (ANE). $\nabla T$, **E**, and **M** denote temperature gradient applied to the sample, the electric field driven by ANE, and a unit vector of the sample magnetization. It demonstrates nanostructure-engineering of ANE in four different scenarios: **a,** Fe-based amorphous sample (Nanomet), **b,** the emergence of nanoprecipitates/clusters in an amorphous matrix, **c,** the high-volume density of nanoprecipitates/clusters embedded in an amorphous matrix, indicating high output signals, and **d,** the crystallized Nanomet samples with reduced ANE signals. The same scenario is applicable to the anomalous Ettingshausen effect (AEE) since ANE is the Onsager reciprocal of AEE.

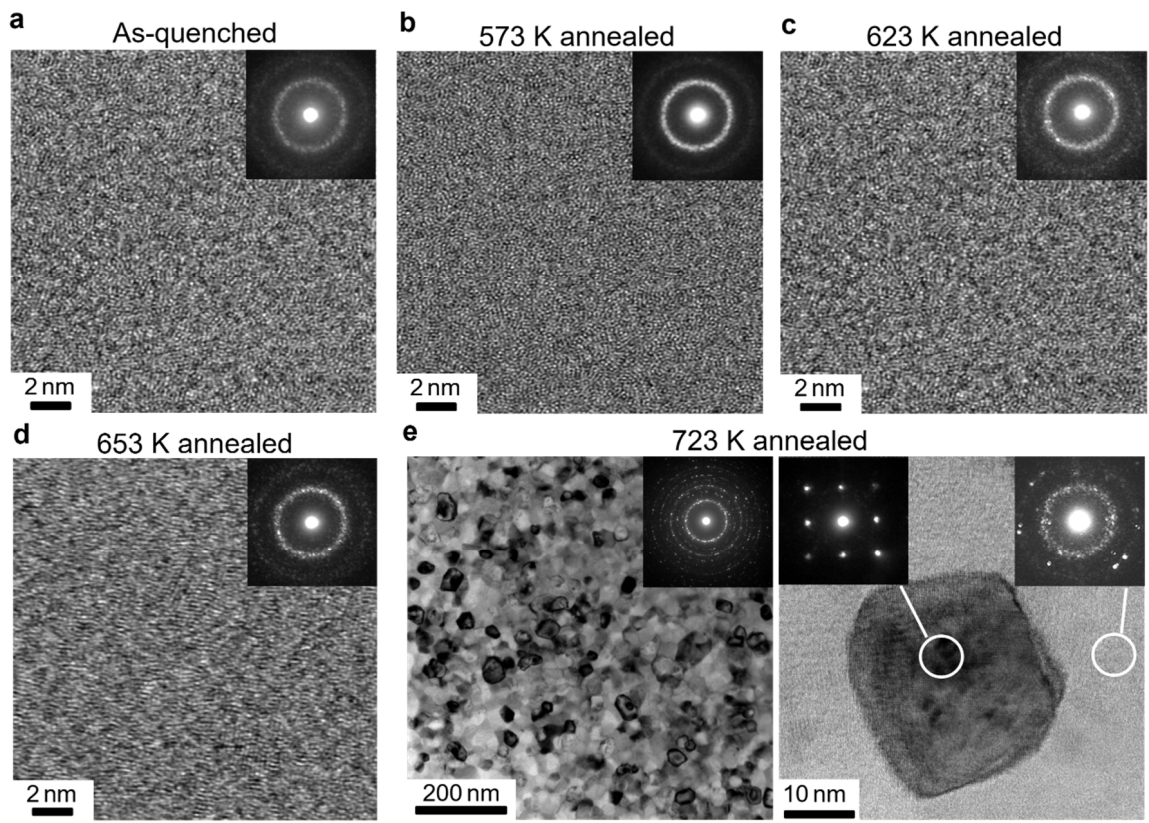

**Fig. 2 | High resolution TEM analysis of annealed Nanomet samples.** High-resolution bright field (BF)-STEM images with corresponding nanobeam electron diffraction patterns obtained from **a,** as-quenched ribbons and post-annealed at temperatures of 573, 623, 653, and 723 K shown in **b-e**, respectively. BF-STEM images of the 723 K annealed sample evidence the crystallization of α-Fe(Si) along with residual amorphous regions.

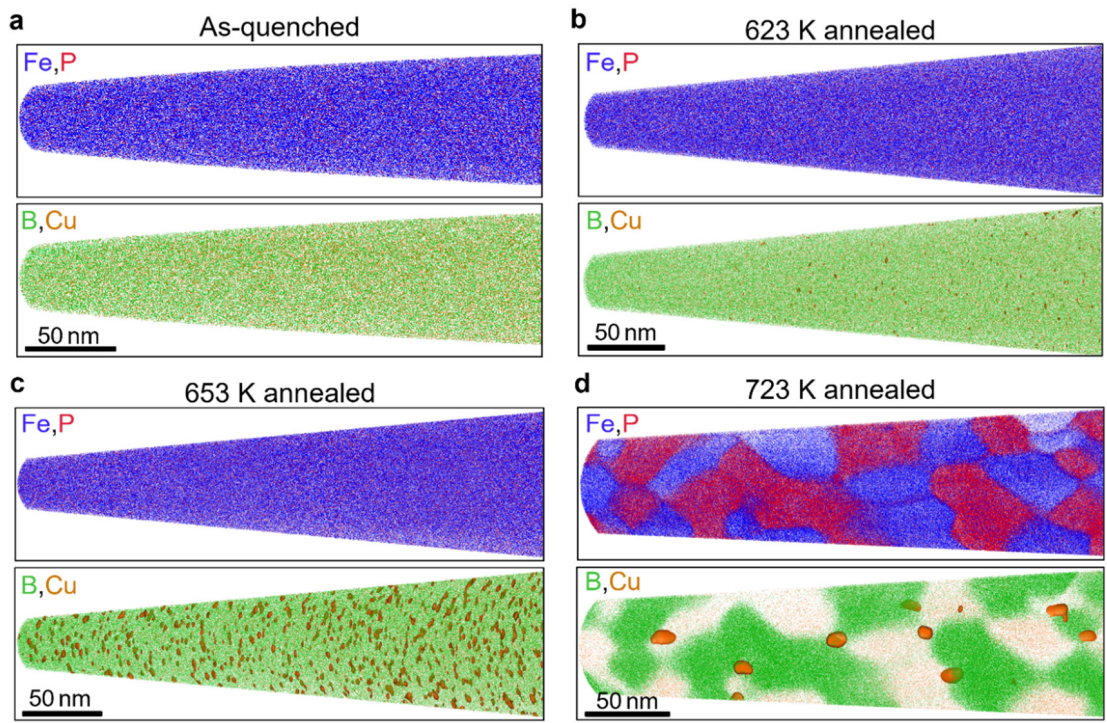

**Fig. 3 | Atom probe tomography analysis of annealed Nanomet samples.** APT elemental maps of Fe (blue), P (red), B (green), and Cu (orange) in different samples: **a**, as-quenched, **b**, annealed at 623 K, **c,** annealed at 653 K and **d,** annealed at 723 K.

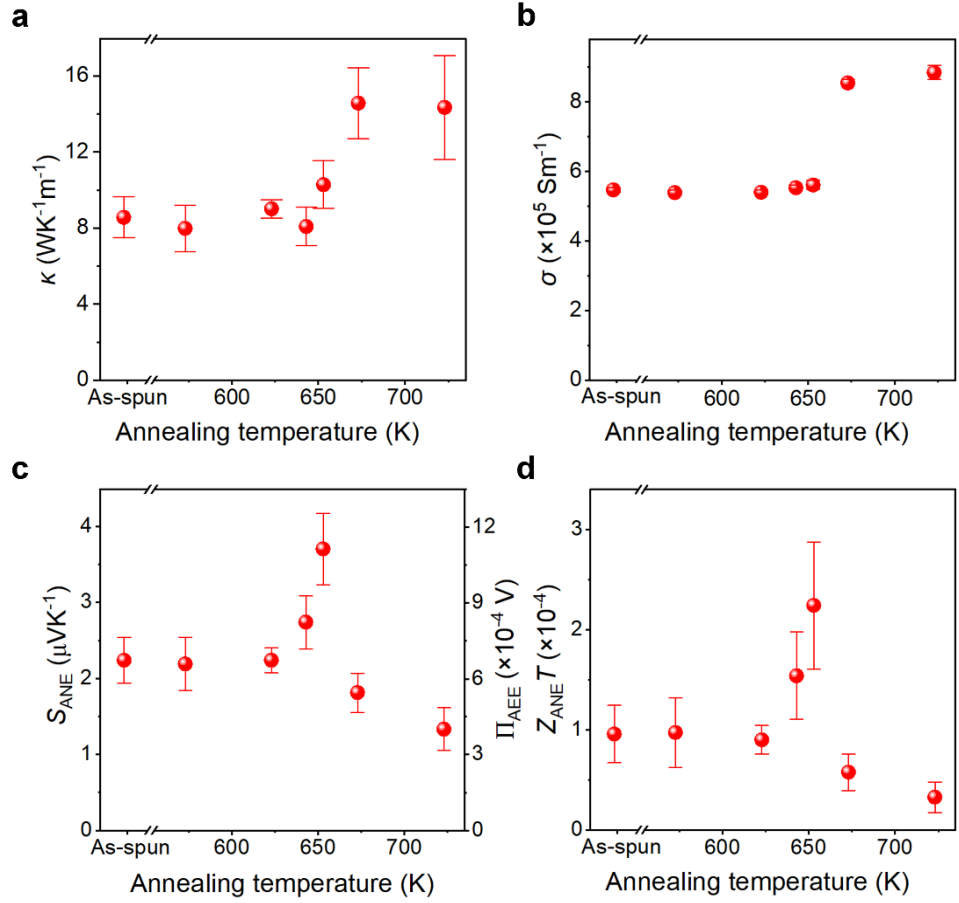

**Fig. 4 | Transport properties of annealed Nanomet samples. a-d,** Annealing temperature dependence of the thermal conductivity $\kappa$ (**a**), electrical conductivity $\sigma$ (**b**), anomalous Nernst coefficient $S_{ANE}$ and the corresponding anomalous Ettingshausen coefficient $\Pi_{AEE}$ estimated using the Onsager reciprocal relation at 300 K (**c**), and dimensionless ANE figure of merit $Z_{ANE}T$ at 300 K (**d**) for the Nanomet samples.

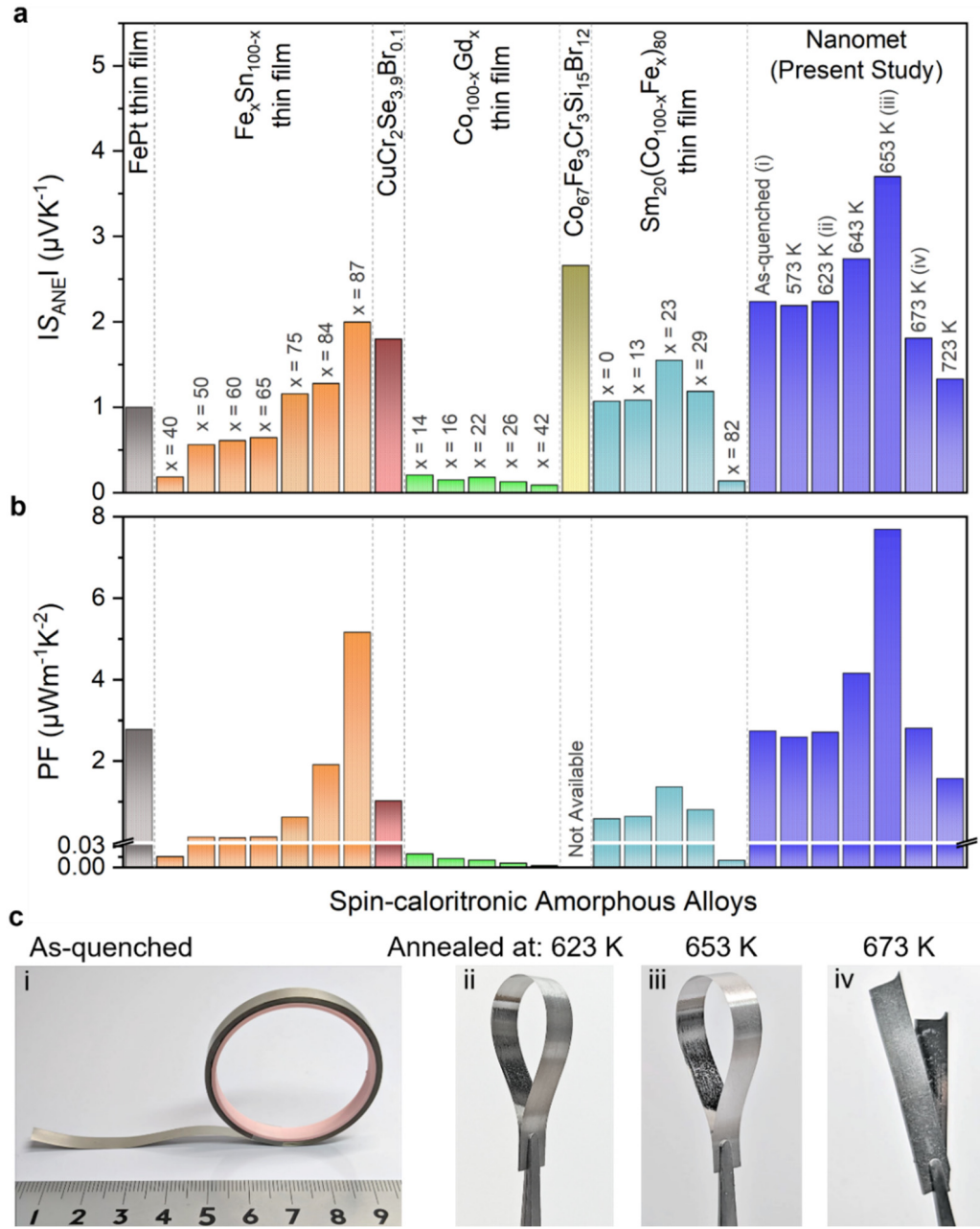

**Fig. 5 | Spin-caloritronic amorphous alloys for the anomalous Nernst effect and demonstration of mechanical flexibility of Nanomet samples.** Comparison of **a,** the absolute values of $S_{ANE}$ and **b,** power factor for various spin-caloritronic amorphous alloys measured at room temperature. Notably, the Nanomet ribbon (present study) annealed at 653 K exhibits the significantly large $|S_{ANE}|$ and power factor value compared to that for the other reported alloys (Fe-Pt[36], $Fe_xSn_{100-x}$[22], $CuCr_2Se_{3.9}Br_{0.1}$[39], $Co_{100-x}Gd_x$[35,38], $Co_{67}Fe_3Cr_3Si_{15}Br_{12}$[37], and $Sm_{20}(Co_{100-x}Fe_x)_{80}$[33]). **c,** Highlighting the mechanical flexibility of long, thin Nanomet ribbons prepared in different conditions: **c.i.** as-quenched, **c.ii.** annealed at 623 K, and **c.iii.,** annealed at 653K. However, **c.iv.** reveals the brittleness of the ribbon annealed at 673 K, as it broke when subjected to bending.

# Supplementary Information

**Thermal Diffusivity**

**Figure S4** illustrates a schematic of the setup consisting of an infrared camera, a diode laser, a function generator, and a LIT system. **Figure S5** shows the $\phi$ images and the corresponding fitted data along the longitudinal axis of the samples ($\theta = 0° - 180°$). The $\phi$ images indicate that clear laser-induced thermal modulation signals appear on the surface of the samples. The $\phi$ distribution has the expected relatively uniform circular shape due to the heat diffusion from the heat point. Furthermore, the fitted experimental data show the linear dependency of $\phi$ with $r$ and good agreement with equation (1) (see Methods).

**Specific Heat**

**Figure S7** shows the specific heat $C_p$ at room temperature of the Nanomet sample with varying annealing temperature measured by DSC. Typically, the amorphous phase has a higher heat capacity than the crystalline phase due to the presence of more vibrational modes in the amorphous material. This leads to a greater energy storage capacity and, therefore, a higher specific heat, which is consistent with the trend seen in the specific heat with annealing temperature.

**Anomalous Nernst/Ettingshausen Effect**

The rectangular Nanomet samples with the dimensions of ~ $10 \times 1 \pm 0.1$ mm$^2$ were used to measure AEE. The sample was fixed on a glass plate with low thermal conductivity to reduce the heat loss from the sample to the sample stage. The two wires (left- and right-side) were connected in series and directed the charge current in opposite direction between the wires. The spatial distribution of the $A_{odd}$ signal is almost uniform in both wires. Although some samples displayed slight variations in $A_{odd}$ values between the left-side and right-side wires due to the difference in wire width, the lock-in amplitude per unit current density, $A_{odd}/j_c$, is the same. The $\phi_{odd}$ value of the left-side (right-side) wire was observed to be ~0° (~180°) with a charge current applied to the $+x$-direction ($-x$ direction) and a magnetic field in the $+y$-direction, satisfying the behaviour of the AEE-induced temperature modulation for samples showing positive $S_{ANE}$ and $\Pi_{AEE}$[1-3]. It can be observed from **Fig. S9** that an apparent current-induced temperature modulation appeared across the entire surface of the ribbons. The $f$ dependence of $A_{odd}/j_c$ showed almost no change, indicating the AEE-induced temperature modulation reaches the steady state in the present $f$ range. The magnitude of $A_{odd}/j_c$ at $f = 1.0$ Hz was used to quantify $\Pi_{AEE}$ and $S_{ANE}$[1].

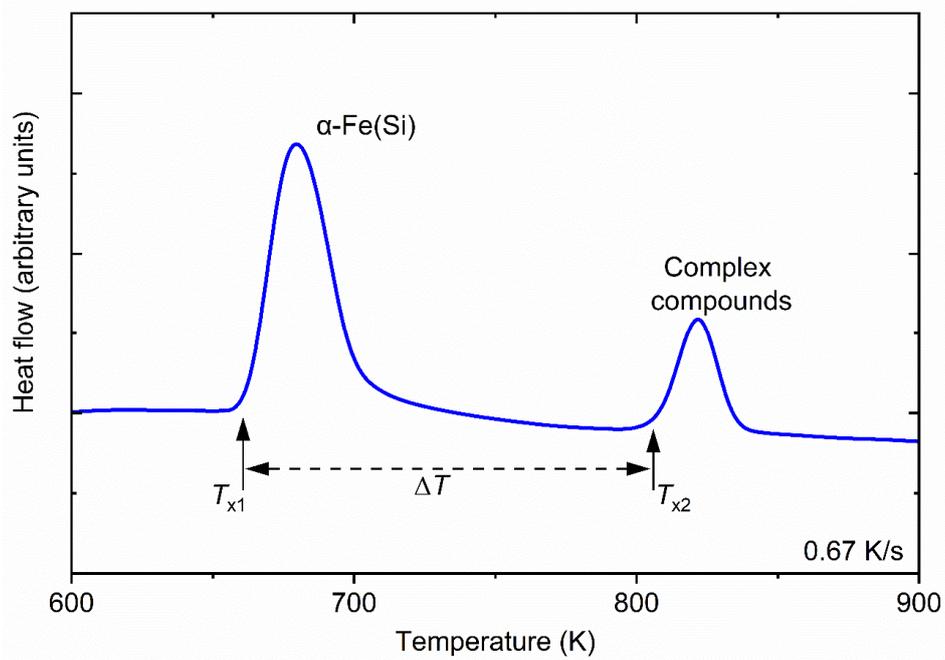

**Fig. S1 | Differential scanning calorimeter (DSC) curve of the as-quenched melt-spun Nanomet ribbon.**

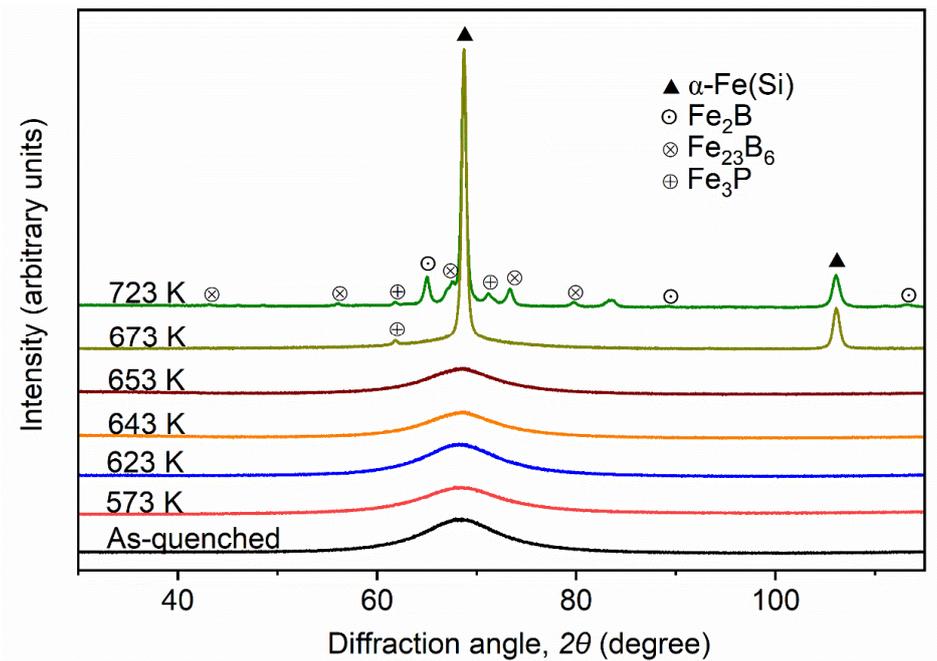

**Fig. S2 | XRD patterns of the annealed Nanomet samples.**

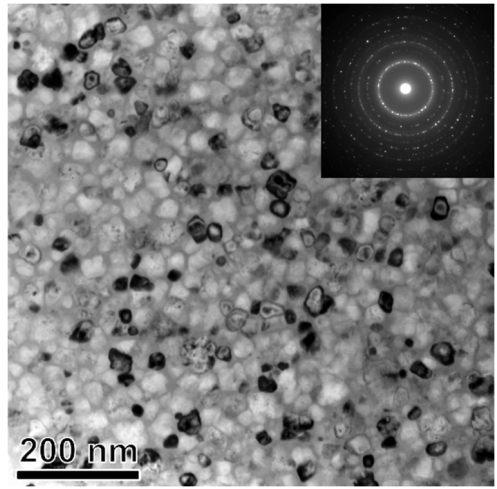

**Fig. S3 | Bright field TEM image of the sample annealed at 643 K with an inset of the selected area electron diffraction pattern.**

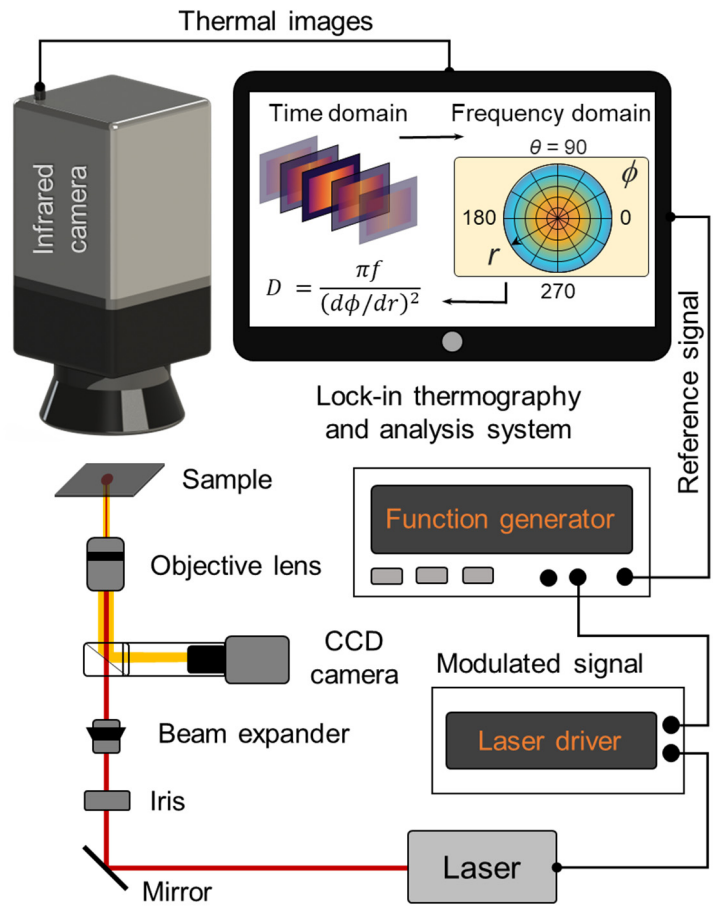

**Fig. S4 | A Schematic of lock-in-thermography (LIT)-based thermal diffusivity measurement technique.** The sample is located in the focal plane of the infrared camera. The diode laser emits a modulated beam at frequency $f$, which is focused on the backside of the sample by the optical setup. The thermal diffusivity $D$ is estimated from the phase $\phi$ correlation to distance $r$ at specific $f$. The angular distribution of $D$ is obtained by revolving the analysis around the laser-heat point at angles $\theta$.

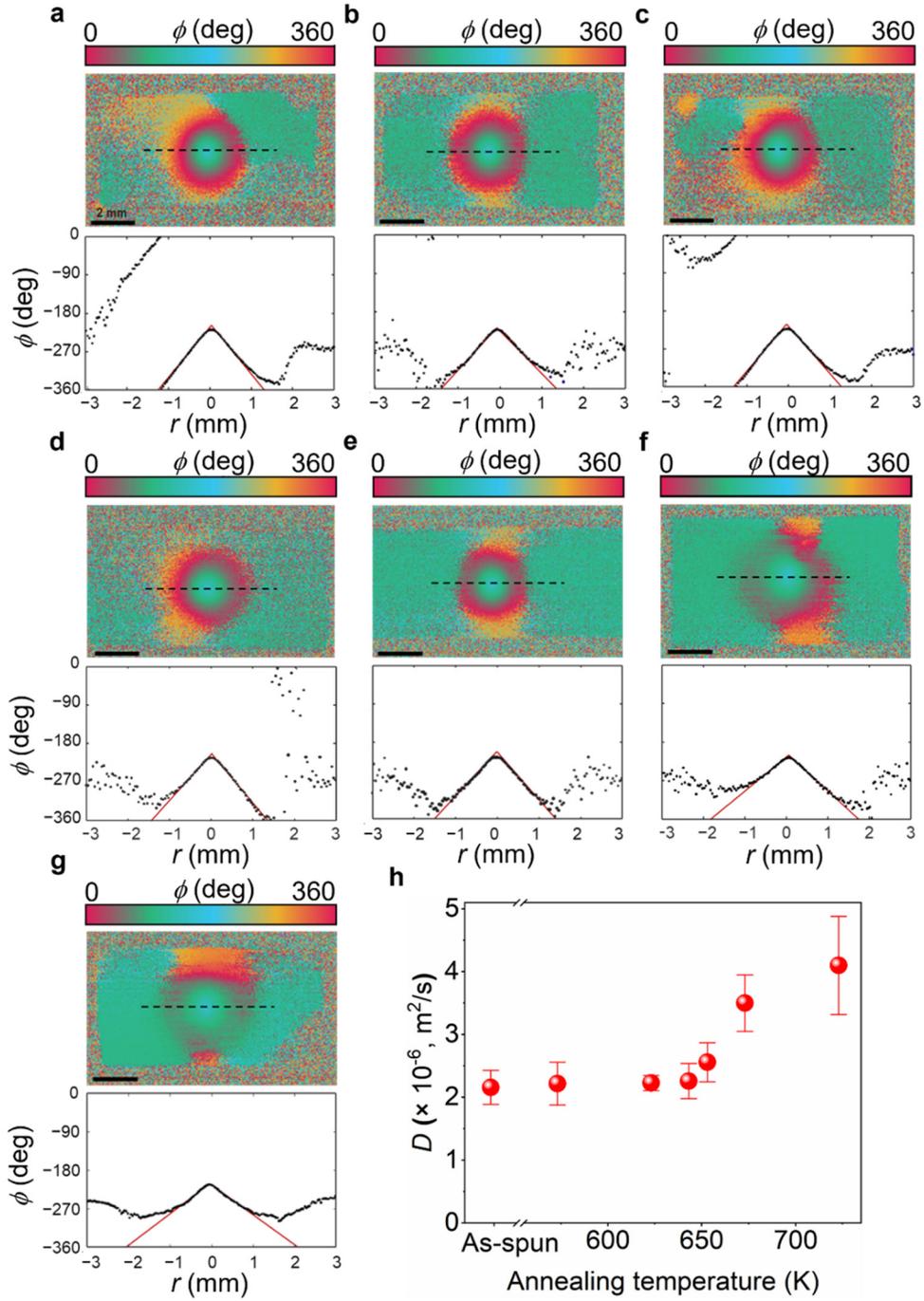

**Fig. S5 | Thermal diffusivity measured of the annealed Nanomet samples.** $\phi$ images with corresponding line profiles along the longitudinal direction (dashed line) of respectively **a,** as-quenched, **b,** annealed at 573 K, **c,** 623 K, **d,** 643 K, **e,** 653 K, **f,** 673 K and **g,** 723 K at $f$ = 3 Hz and laser power $P$ = 10 mW. $r$ = 0 in the line profiles was determined by the position of the laser heating. The solid red lines in **a-g** represent the fitting results using $D = \pi f/(d\phi/dr)^2$. **h,** $D$ values of the annealed Nanomet samples.

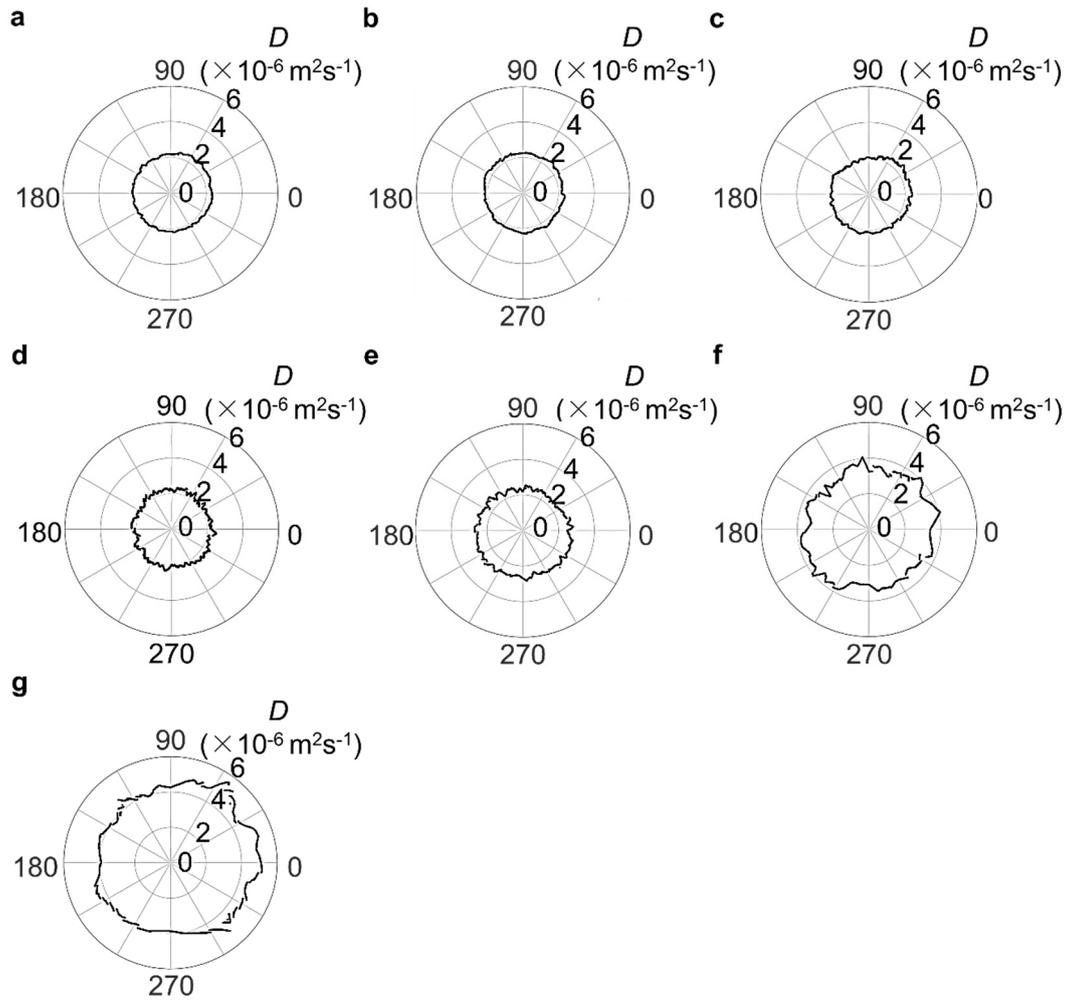

**Fig. S6 | Investigation of angular distribution of thermal diffusivity for the annealed Nanomet samples.** Angular distribution of $D$ ($\times 10^{-6}$ m$^2$s$^{-1}$) measured at $f = 3$ Hz and $T = 300$ K for **a,** as-quenched sample, **b,** annealed at 573 K, **c,** 623 K, **d,** 653 K, **e,** 673 K, and **f,** 723 K. The centre of the graph is located on the position of the laser heating while 0°-180° and 90°-270° axis represents respectively the longitudinal and the transverse axes of the sample.

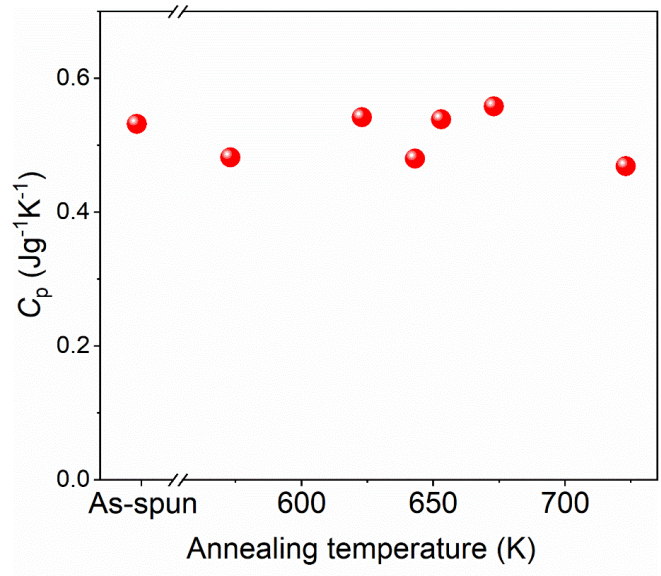

**Fig. S7 | Specific heat $C_p$ as a function of annealing temperature for the Nanomet samples.**

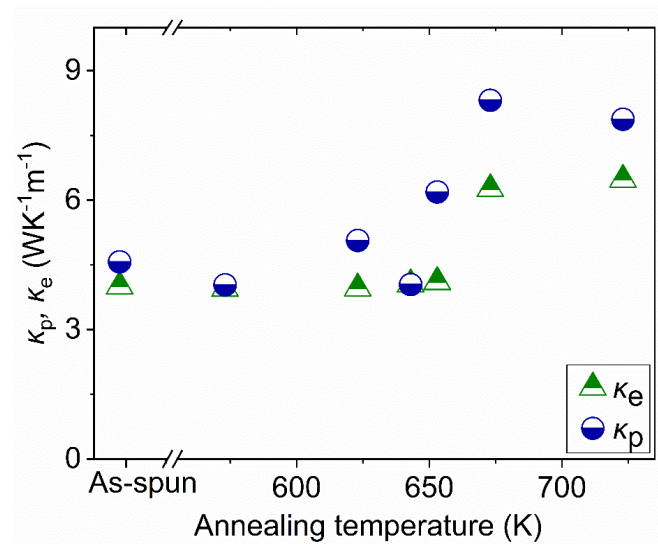

**Fig. S8 | Phonon thermal conductivity $\kappa_p$ and electron thermal conductivity $\kappa_e$ as a function of annealing temperature for the Nanomet samples.**

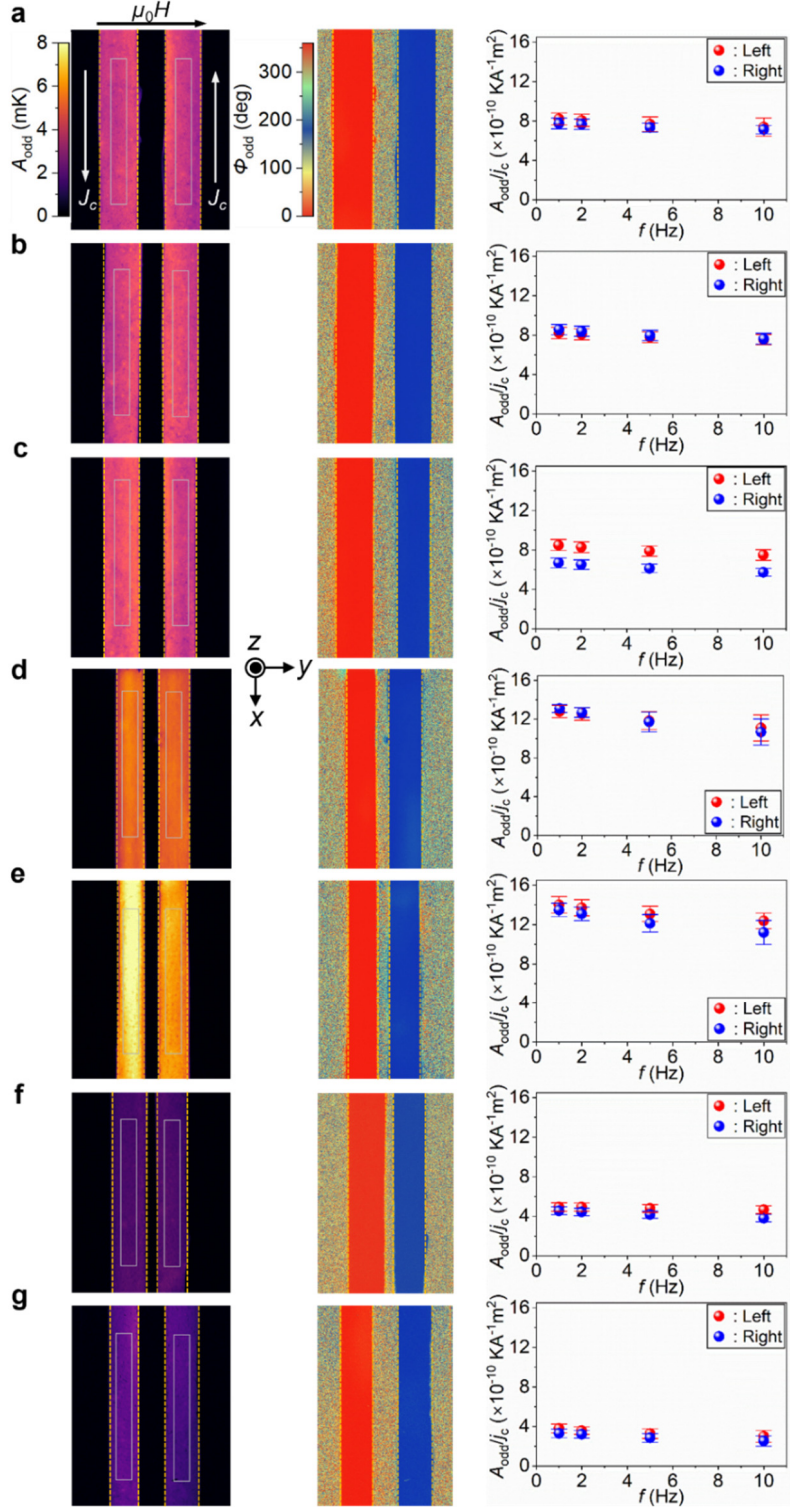

**Fig. S9 | Lock-in thermography measurement of AEE for the annealed Nanomet samples.** Field-odd component of lock-in amplitude ($A_{odd}$) and phase ($\phi_{odd}$) images at $\mu_0 H$ = 1.0 T and the corresponding frequency $f$ dependence of $A_{odd}/j_c$ for **a,** as-quenched, **b,** annealed at 573 K, **c,** 623 K, **d,** 643 K, **e,** 653 K, **f,** 673 K and **g,** 723 K samples. $J_c$ denotes the charge current applied to the sample with charge current density $j_c$. The values and error bars in the rightmost graphs represent the average and standard deviation of $A_{odd}$ on the plot area shown by grey rectangular boxes in the $A_{odd}$ images, respectively.